\documentclass[prl,twocolumn,showpacs]{revtex4}
\usepackage{graphicx}
\begin{document}
\title{Avalanches in a Bose-Einstein Condensate}
\author{J. Schuster, A. Marte, S. Amtage, B. Sang and G. Rempe}
\address{Max-Planck-Institut f\"{u}r Quantenoptik, Hans-Kopfermann-Str. 1,
D-85748 Garching, Germany}
\author{H. C. W. Beijerinck}
\address{Physics Department, Eindhoven University of Technology, P.O. Box 513,
        5600 MB Eindhoven, The Netherlands}
%\date{\today}
\begin{abstract}
Collisional avalanches are identified to be responsible for an $8$-fold increase
of the initial loss rate of a large $^{87}$Rb condensate. We show that the
collisional opacity of an ultra-cold gas exhibits a critical value. When
exceeded, losses due to inelastic collisions are substantially enhanced.
Under these circumstances, reaching the hydrodynamic regime in conventional BEC
experiments is highly questionable.
\end{abstract}
\maketitle
One of the current goals in the field of Bose-Einstein
condensation (BEC) is the production of a condensate in the
collisionally opaque or hydrodynamic regime, where the mean free
path of an atom is much less than the size of the sample. This
would offer the opportunity to study striking phenomena like
quantum depletion or dynamical local thermal equilibrium. In this
context, one possible approach is to increase the interaction
among the atoms by means of Feshbach resonances \cite{FESHBACH}.
It has been observed, however, that in their vicinity the large
cross-section for elastic collisions is accompanied by a dramatic
increase of atom losses \cite{STE99,ROB00}. Hence, it seems
advantageous to follow a different route by producing large and
dense condensates.

In this letter we conclude that the collisionally opaque regime can hardly be
reached in alkali BEC experiments. We identify an intrinsic decay process that
severely limits the average column density $\langle nl\rangle$ of condensates
at values achieved in present BEC experiments. It is based on collisional
avalanches that are triggered by inelastic collisions between condensate atoms.
A considerable part of the energy released in these initiatory collisions is
distributed among trapped atoms resulting in a dramatic enhancement of the
total loss from the condensate. In analogy to the critical mass needed for a
nuclear explosion, we define a critical value of the collisional opacity
$\langle n l \rangle \sigma_s$, with $\sigma_s = 8 \pi a^2$ the s-wave cross
section for like atoms and $a$ the scattering length. The {\em critical opacity}
equals 0.693, corresponding to a collision probability of 0.5. Related scenarios
have been discussed in Refs. \cite{COR99,GUE99}, but were assumed to play a
minor role in the experimentally relevant region. However, we present strong
experimental evidence that the anomalous decay of our $^{87}$Rb condensate is
caused by collisional avalanches. This is supported by the good agreement of a
simple model with the data.

The crucial point for the occurrence of an avalanche is whether
the products of a one-, two- or three-body decay process have a
substantial probability
%------------------------------------------------------------------------------
\begin{eqnarray}
p(E) &=& 1- \exp\left[-\langle nl\rangle\, \sigma (E)\right],
\label{eq:probability}
\end{eqnarray}
%------------------------------------------------------------------------------
of undergoing secondary collisions before leaving the trap
\cite{BEI00b}, with $\sigma (E)$ the total cross section at
kinetic energy $E$. The collision probability varies significantly
with temperature and is usually highest in the s-wave scattering
regime.
%%%%%%%%%%%%%%%%%%%%%%%%%%%FIGURE 1%%%%%%%%%%%%%%%%%%%%%%%%%%%%%%%%%%%%%%%%%%%%
\begin{figure} 
\includegraphics{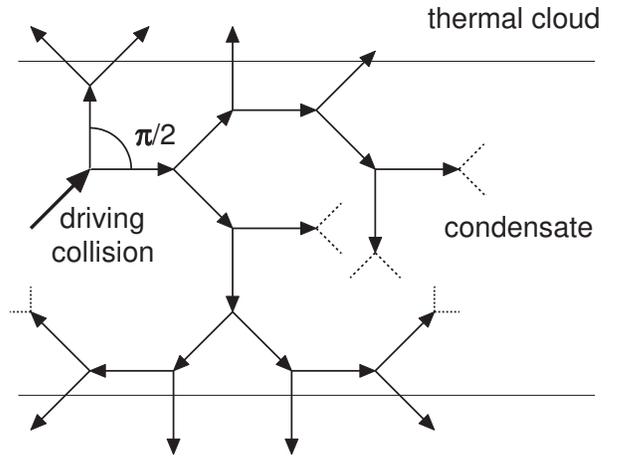}
\caption{Sketch of a collisional avalanche in a homogeneous
condensate with a radius equal to twice the mean free path $(n \sigma_s)^{-1}$,
illustrating the enhancement of the loss rate. \label{fig:figure1}}
\end{figure}
%%%%%%%%%%%%%%%%%%%%%%%%%%%FIGURE 1%%%%%%%%%%%%%%%%%%%%%%%%%%%%%%%%%%%%%%%%%%%%
Here, the differential cross section is isotropic in the center-of-mass system;
in the lab system the two atoms fly apart at an angle of $\pi/2$ on average.
The energy of the projectile is on average equally distributed among the two
colliding atoms. This implies that each collision results in {\em two} new atoms
that both can continue their collisional havoc in the trap until they leave the
condensate (Fig. \ref{fig:figure1}). If the probability for collisions is higher
than 0.5, the average number of colliding atoms increases with every step
of the collisional chain which now becomes self-sustaining.

To calculate the total loss from the condensate, we start from the
well known loss rates $\dot N_i = - K_i N \langle n^{i-1}\rangle$,
with $i=1,2,3$ associated with one-, two- and three-body decay
processes with rate constant  $K_{i}$, respectively. Here, $N$ is
the number of atoms in the gas with the density distribution
$n(\vec r)$. Depending on the energy of the decay products,
typically a few or even no further collisions are needed to
generate an atom with an energy $E_{i,s}$ whose next collision
would be in the s-wave regime. The probability for this
collisional chain is $p_{i,1}\cdot p_{i,2}\cdots = \tilde p_{i}$
with $p_{i,n}=p(E_{i,n})$. During this process, on average $\tilde
g_i$ atoms are lost from the condensate without undergoing
secondary collisions. Next, an atom with energy $E_{i,s}$ induces
an avalanche with a collision probability $p_{s}=1-\exp [-\langle
nl\rangle \sigma _s]$ that now is independent of energy.
Consequently, atoms with an energy $E_{i,s}/2^k$ in the
$k^{\mbox{\small th}}$ step of an avalanche are generated with a
probability of $\tilde{p_{i}}p_s^k$. Since every collision now
results in two projectiles in the next step, the degeneracy of
step $k$ is $2^k$. The rate at which atoms are lost from this
avalanche step is $\tilde{p_{i}}\dot{N_{i}}2^kp_s^k(1-p_s)$ and
the total loss rate from the condensate becomes
%------------------------------------------------------------------------------
\begin{eqnarray}
\dot{\cal N}_{i,aval} &=& \dot{N_{i}}\left[\tilde g_i+\tilde
p_{i}(1-p_{s})\sum_{k=0}^{k^{max}_{i}}
               2^{k}p_{s}^{k}\right]\,,
\label{eq:aval}
\end{eqnarray}
%------------------------------------------------------------------------------
where the sum extends over all  relevant avalanche steps.

To determine the cutoff $k_{i}^{max}$, note that
avalanche-enhanced losses can occur up to a step with an energy on
the order of the chemical potential. However, when the energy
falls short of the trap depth, $E_{trap}$, the atoms lost from the
cold sample are still trapped. They will repeatedly penetrate the
cloud and thus give rise to heating. This will either be
compensated by an evaporation of atoms from the trap or will
reduce the condensed fraction by increasing the temperature. Both
possibilities are not described by Eq. (\ref{eq:aval}). Therefore,
we use $k_{i}^{max}\leq \log_{2}(E_{i,s}/E_{trap})$ as a cutoff
and, hence, account only for immediate trap losses.

The additional heat-induced depletion caused by trapped avalanche
atoms can easily be estimated for the case that the temperature is
fixed by the trap depth. Each atom participating in step
$(k^{max}_{i}+1)$ of the avalanche will finally dump about the
energy $E_{trap}/2$ into the system. Since any evaporated atom
takes the energy $E_{trap}$ with it, about half as many atoms as
are produced in the step $(k^{max}_{i}+1)$ of the avalanche have
to be evaporated to keep the temperature constant. Hence, the
evaporation rate is
%-----------------------------------------------------------------------------
\begin{eqnarray}
\dot{\cal N}_{i,heat} \simeq ({1}/{2})\,\tilde p_{i}\dot
N_{i}\,2^{(k_{i}^{max}+1)}p_{s}^{(k_{i}^{max}+1)}\,.
\label{eq:heating}
\end{eqnarray}
%-----------------------------------------------------------------------------

Equation (\ref{eq:aval}) predicts substantially enhanced losses as soon as the
critical opacity is exceeded. However, for a given $k_{i}^{max}$ there is a
second critical value of the opacity above which the loss rate
$\dot{\cal N}_{i,aval}$ decreases again. Now, with increasing opacity the
limited trap depth continuously looses its shielding effect against the
products of inelastic collisions, since most avalanches generate
trapped particles. In the collisional regime with $p_s\simeq 1$, the energy
released in an inelastic process will be entirely dissipated in the system.
This results in an explosion-like particle loss according to Eq.
(\ref{eq:heating}).

To apply our model, the column density must be evaluated according to
%------------------------------------------------------------------------------
\begin{eqnarray}
\langle nl\rangle &=& \int [n(\vec r\,)/N]
        \cdot \int [n(\vec r+\vec R\,)/4\pi R^2]\,d^3 R\,d^3 r \,\\
\label{eq:coldens}
        &=&c\cdot n_pW_{\perp}\,\int_{0}^{\infty}
        dx\,(1+x^2)^{-1}\,(1+\varepsilon^2 x^2)^{-1/2} ,
\label{eq:coldensnum}
\end{eqnarray}
%------------------------------------------------------------------------------
where the second line is the result for the case of a harmonic potential with
cylindrical symmetry. Here, $\varepsilon =\omega_{\parallel}/\omega_{\perp}$ is
the ratio of the trap frequencies and $n_p$ is the peak density of the cold
sample. For the parabolic density distribution of the condensate, $W_{\perp}$
is the half radial width and $c=5/12$. Due to the scaling $n_p\propto N^{2/5}$
and $W_{\perp}\propto N^{1/5}$ the column density of a condensate scales as
$N^{3/5}$, so that the effect of multiple collisions is quite persistent. For
a Gaussian distribution we find $W_{\perp}=\sigma_{\perp}$, $c=\sqrt{\pi/8}$
and a scaling according to $\langle nl\rangle\propto N$. In a harmonic
potential, the ideal Bose distribution can be represented as a sum of Gaussian
distributions and the latter result can thus be used to evaluate the column
density close to degeneracy. For a Bose distribution the opacity scales
disproportionate to $N$, resulting in a faster decline of the avalanche
enhancement.

The next step is to identify the energies of the initial decay
products. For a background gas collision, $E_{1,1}$ depends on the
mass of the impinging particle that is assumed to be Rb in our
system. In the case of spin relaxation, $E_{2,1}$ equals either
the Zeeman energy or the hyperfine splitting energy. For
three-body recombination $E_{3,1}$ has to be derived from the
binding energy of the most weakly bound level in which the dimer
is predominantly formed. Clearly, the molecule is likely to be
deactivated in a subsequent inelastic collision with a condensate
atom \cite{YUR99,WYN00}. Deactivating collisions will be a serious
problem in highly opaque clouds where atoms with higher energies
still have high collision probabilities. In our experiment,
however, the collision probability is significantly smaller at
typical deactivation energies of 0.1 K than at the binding energy
of the molecules in the last bound level. In our analysis we
therefore do not account for avalanches triggered by deactivating
collisions. The values of all parameters used to calculate the
effective losses are listed in Table \ref{tab:table1}. Note that
in order to account for the avalanches triggered by the two-body
decay, the partial rates associated with the various exit channels
are needed since they correspond to different energies released in
the process.

%
%%%%%%%%%%%%%%%%%%%%%%%%%%%%%%%%%%%% TABLE 1 %%%%%%%%%%%%%%%%%%%%%%%%%%%%%%%%%
\begin{table}[bt]
\caption{Rate constants for the initiatory processes and energies
of the subsequent collisions that are necessary to generate the
first avalanche atom with energy $E_{s}$.}
\begin{tabular}{llcc}
$i$ & type& {rate constant} & {$\Delta E_1,\ldots,\Delta E_n=E_s$}\\
    &     & {$K_i$} & {[(3/2)$k_{B}$]} \\
 \hline 1 & background  & {$1/(39\,s)^{a}$}  & 4\,K, 100, 5, 0.5\,mK \\
 \hline 2& Zeeman$\,^{b}$           & $1.4\times 10^{-18}\,$cm$^3$/s  & 0.022\,mK\\
         & 2$\times$Zeeman$\,^{c}$ & $3.7\times 10^{-17}\,$ cm$^3$/s & 0.045\,mK\\
         & hyperfine$\,^{d}$        & $2.2\times 10^{-16}\,$ cm$^3$/s & 109, 5, 0.5\,mK\\
         & 2$\times$hyperfine$\,^{e}$& $1.3\times 10^{-16}\,$ cm$^3$/s & 219, 8, 0.5\,mK\\
 \hline 3 & recombination$^{f}$ & $1.8\times 10^{-29}\,$cm$^6$/s  & 0.54\,mK$\,^{g}$
\end{tabular}
$^a$\cite{TAU}; $^{b,c,d,e}$\cite{VERpca,VERpcb}; $^f$\cite{SOE99,ESR99}; $^g$\cite{HEIpc};
\label{tab:table1}
\end{table}
%%%%%%%%%%%%%%%%%%%%%%%%%%%%%%%%%%%% END TABLE 1 %%%%%%%%%%%%%%%%%%%%%%%%%%%%%%
%

Finally, the presence of a diffuse atom cloud in the trapping volume can cause
additional losses (see e.g. \cite{COR99}). In a steep magnetic trap with a
depth of a few mK, such an "Oort" cloud is mainly a consequence of incomplete
evaporation at high magnetic fields \cite{DES99} or low radio frequency (rf)
power. In our experiment, the temperature of the diffuse cloud will probably
be on the order of 400 $\mu$K, corresponding to the measured initial
temperature of the magnetically trapped cloud. Even in a rf-shielded trap
these atoms will penetrate the condensate giving rise to an additional decay
rate according to $1/\tau=n_{\rm oort}\sigma_sv_{\rm oort}$, with
$n_{\rm oort}$ and $v_{\rm oort}$ the density and the thermal velocity of the
penetrating atoms, respectively. Collisions with Oort atoms will also trigger
avalanches, because the collision energy is close to the s-wave regime.

To compare our data with the predictions of the model, the differential and
the total scattering cross sections are needed. Above a kinetic energy
$E/k_B$ of $60$ mK we calculate the energy transfer by collisions using a
model function for the small-angle differential cross section \cite{BEI80}.
For collisions below $60$ mK we use the numerical results from a full
quantum treatment \cite{VERpca}. For $^{87}$Rb in the $|2,\,2\rangle$ state,
the large contribution of a d-wave scattering resonance to the total cross
section leads to $\sigma (E)\simeq 4\times\sigma _s$ at an energy of
$E/k_B=560\,\mu$K in the lab system. This almost exactly coincides with the
energy transferred to the third atom in a recombination event (table
\ref{tab:table1}). Hence, a secondary collision of this atom will occur with
a probability of 0.99 already when the probability for s-wave collisions
among condensate atoms is 0.7. For kinetic energies $E/k_B \leq 1.5$ mK,
the total cross section obeys $\sigma (E)\gtrsim \sigma _s$. For simplicity,
we use $\sigma _s$ for calculating avalanches in this energy range. Our
model therefore yields a lower bound for the total loss.

The apparatus used to study the condensates has been described
previously \cite{ERN98,ERN98a}. The experiment is performed with
$^{87}$Rb atoms in the $|2,\,2\rangle$ state. A Ioffe-Pritchard
magnetic trap with a bias field of 2 G and oscillation frequencies
of $\omega_{\perp}/2\pi = 227$ Hz and $\omega_{\parallel}/2\pi =
24.5$ Hz is used. The atoms are cooled by rf evaporation and then
held in the trap for a variable time interval. During the storage
time, the trap depth is set to $E_{trap}/k_B = 4.4$ $\mu$K by
means of the rf shield. From the width of the density distribution
after expansion the atom number $N_C$ in the condensate is
determined. At minimum storage time, we find $N_C = 1.1\times10^6$
atoms and $n_p = 6.4\times 10^{14}$ cm$^{-3}$ and no discernible
non-condensed fraction.

\begin{figure}
\includegraphics{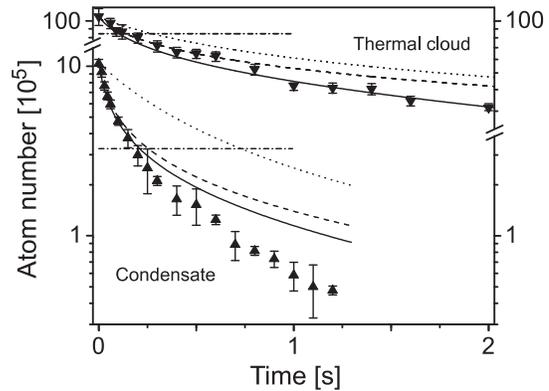}
\caption{Decay of the
condensate and the thermal cloud. The horizontal line corresponds to the
critical opacity. For comparison, the calculated decay due to the initial
one-, two- and three-body loss rates without (dotted line) and with
avalanche-enhancement (dashed line) are shown. The full line includes the
effect of an Oort cloud with avalanche enhancement. \label{fig:figure2}}
\end{figure}

The decay curve of the condensate is shown in Fig.
\ref{fig:figure2}, revealing that about half the initial number of
atoms is lost within the first 100 ms. The dotted line shows the
theoretical prediction assuming that losses occur solely due to
background gas collisions, spin relaxation and recombination
(Table \ref{tab:table1}). The observed loss is 8 times faster than
predicted. Moreover, the additional decay is clearly
non-exponential and can therefore not result from primary
collisions with Oort atoms. Hence, multiple collisions have to be
taken into account.

Indeed, with $\langle nl\rangle \sigma _s=1.4$ the critical
opacity is considerably exceeded. To the best of our knowledge,
the corresponding s-wave collision probability of $p_s=0.76$ has
not been reached in published work on Rb condensates in the
off-resonant scattering regime. This explains why our observations
differ from those made in other experiments
\cite{BUR97,SOE99,DIEpc}. The dashed line displayed in Fig.
\ref{fig:figure2} has been obtained by numerically integrating the
rate equation $\dot{N} =\sum_{i=1}^3(\dot{\cal
N}_{i,aval}+\dot{\cal N}_{i,heat})$ that describes
avalanche-enhanced losses according to Eqs. (\ref{eq:aval}) and
(\ref{eq:heating}), without any adjustable parameter and
neglecting the contribution of an Oort cloud. We find good
agreement between theory and experiment within the first 200 ms,
showing that collisional avalanches triggered by recombination
events are responsible for the fast initial decay.
%%%%%%%%%%%%%%%%%%%%%%%%%%%FIGURE 2%%%%%%%%%%%%%%%%%%%%%%%%%%%%%%%%%%%%
%%%%%%%%%%%%%%%%%%%%%%%%%%%FIGURE 2%%%%%%%%%%%%%%%%%%%%%%%%%%%%%%%%%%%%%

To investigate the role of an Oort cloud, we have performed a similar
experiment with an atom cloud at a lower density. Figure \ref{fig:figure2}
shows the decay of a non-condensed cloud with $1\times 10^7$ atoms at a
temperature of 1 $\mu$K and a peak density of $3.5\times 10^{14}$ cm$^{-3}$.
The number of atoms is determined from the total absorption of near-resonant
laser light. The trap depth is limited to $10\,\mu$K, according to
the higher temperature of the sample. Again, the decay is non-exponential and
initially about two times faster than predicted by the primary loss-rates
(dotted line). At an opacity of 0.9, obtained by assuming an ideal
Bose distribution, we already expect a weak avalanche-enhancement. This
allows us to test our model in a different regime since in a thermal cloud
avalanches are less persistent than in a condensate. In addition, the
intrinsic two- and three-body decay rates will die out during the
observation time whereas the effect of an Oort cloud as a one-body decay will
persist. The solid line in Fig. \ref{fig:figure2} is the prediction of our
model where we have included an avalanche-enhanced decay rate caused by an
Oort cloud. Good agreement with the data is obtained for
$1/\tau=1/7.8\,{\rm s}$, corresponding to $n_{\rm oort}=5\times10^8$
cm$^{-3}$ at 400 $\mu$K. Such a density is produced by only a few times
$10^{5}$ atoms and appears realistic in view of the more than $10^{9}$ atoms
that were loaded into the magnetic trap. It is also consistent with the fact
that we have no direct experimental evidence for an Oort cloud and that the
initial decay is correctly predicted by the model even if the contribution
of the Oort cloud is neglected (dashed line).

We can now calculate the extra loss rate of the condensate due to an Oort
cloud. Since the two experiments described above are performed under
identical conditions, the density of the Oort cloud is essentially
unchanged in the two measurements. As can be seen from the solid
line in Fig. \ref{fig:figure2}, the small extra contribution from the
Oort cloud does not change the predicted initial decay but slightly
improves the agreement between the model and the data for longer times.
The small remaining discrepancy can be the result of an additional decay
not accounted for in our model. In particular, avalanches will seriously
perturb the equilibrium of the condensate by inducing local fluctuations
of the mean-field energy \cite{GUE99}. Since the damping rate of
excitations can be small compared to the elastic collision rate, we expect
that this process introduces a second time scale to the decay that depends
on the history of the condensate.

The simultaneous agreement of our model with the two complementary data sets
strongly supports the evidence for the occurrence of collisional avalanches
in our experiments. Our analysis reveals that the density of a cold gas is
severely limited as soon as the s-wave collisional opacity exceeds the
critical value of 0.693. It is important to point out that the anomalous
initial decay of the condensate is attributed to collisional avalanches
almost exclusively triggered by the intrinsic process of recombination and
that no free parameters are introduced in the model. We have no evidence for
the contribution of an Oort cloud to the fast initial decay observed in our
experiments.

We conclude that it will be hard to enter the collisional regime in alkali BEC
systems. For $^{87}$Rb in the $|2,\,2\rangle$ state the prospects are even
worse due to the large collision cross section of the recombination products.
Hydrodynamic conditions might be reached in the longitudinal direction in an
extremely prolate geometry, as can be seen from Eq. (\ref{eq:coldensnum}).
Avalanches triggered by recombination events can be suppressed in the vicinity
of Feshbach resonances, where the recombination energy becomes very small.
However, collisional deactivation of the highly excited molecules can still
produce avalanches \cite{YUR99} and might contribute to the fast decay reported
in Ref. \cite{STE99}. This offers a new application for a condensate of, e.g.,
ground-state helium atoms, where recombination is not possible.

The authors are indebted to B. J. Verhaar for providing us with
the results of calculations regarding the scattering cross
sections and the spin relaxation rates. H.C.W.B. gratefully
acknowledges the hospitality of JILA, NIST and the University of
Colorado, Boulder.

\end{document}